\documentclass[journal]{IEEEtran}

\usepackage{graphicx}
\usepackage{commath}
\usepackage{float}
\usepackage{lipsum}
\usepackage{textcomp}
\usepackage{amsmath, esint}
\usepackage{amssymb}
\usepackage{mathtools}
\usepackage{siunitx}
\usepackage{comment}
\usepackage{array}
\usepackage{threeparttable}
\usepackage{stackengine}
\graphicspath{ {./images/} }
\usepackage{tabularx}
\usepackage{xcolor}
\usepackage{tablefootnote}
\usepackage{tabularx}

\newcolumntype{P}[1]{>{\centering\arraybackslash}p{#1}}
\newcolumntype{M}[1]{>{\centering\arraybackslash}m{#1}}

\begin{document}

\title{Flexible Phased Array Sheets: A Techno-Economic Analysis}

\author{Oren S. Mizrahi, Austin Fikes, Ali Hajimiri
\thanks{O.S.M., A.F., and A.H. are with the Department
of Electrical Engineering, California Institute of Technology, Pasadena,
CA, 91125 USA.}
\thanks{Manuscript received February 7, 2023}}

\markboth{ }{}

\maketitle

\begin{abstract}
Phased arrays have enabled advances in communications, sensing, imaging, and wireless power transfer. In all these applications, large apertures enable higher power, higher data rates, higher resolution, and complex functionalities, but are elusive owing to a correspondingly large cost, mass, and physical size. Flexible phased arrays (FPAs) show potential in breaking this trade-off. Their thinness and extremely low mass allow FPAs to be folded, rolled, or otherwise compressed into smaller sizes, thus enabling new regimes of transport and entirely new applications currently not possible. Though a number laboratory prototypes of FPAs have been constructed, the economics of large-scale FPA production has yet to be explored. This paper presents a model FPA architecture and a cost model for producing it at large-scale. The estimate of the per-unit-area cost is bounded by a three-tiered approach. The cost model projects a ``middle" estimate for FPA production at \$89 per square meter. Estimates for aerial mass density and startup cost are also discussed. This cost model demonstrates that an FPA can be produced at an efficient price point and can potentially replace existing solutions for space, communications, and vehicular applications that demand lightweight, portability, and durability in extreme conditions.
\end{abstract}

\section{Introduction}
Phased arrays based on radio frequency integrated circuits (RFICs), introduced two decades ago, have spurred ongoing inquiry both in academia and industry. This explosion in interest is owing to RFICs' incredible capabilities and diverse applications. By packaging all of the frequency generation, power amplification, and advanced digital control functionalities into a single, extremely thin monolithic platform only several square millimeters in area, integrated circuits enable the large scale and low cost that have allowed phased arrays to find fertile ground in a plethora of diverse applications. IC-driven phased arrays have found utility in communications \cite{YangMTT,Natarajan77GHzTX2006}, imaging \cite{FatemiOpt}, sensing \cite{McIntoshGRS,BrautigamGRS}, and wireless power transfer (WPT) \cite{MatanIMS2020,matan2022npj}.

Previously limited by niche technology nodes, RFICs are now commonly fabricated in CMOS, allowing for incredibly low-cost and highly scalable technology platforms that can also leverage decades of advanced digital design legacy. Flexible phased arrays (FPAs) are a natural extension to this progression, facilitating all the aforementioned applications in novel physical contexts and potentially at lower cost. FPAs that are thin, low-mass, durable, low-cost, and driven by integrated electronics have been proposed for use in communication \cite{IMS_2022_A, IMS_2022_B}, wireless power transfer \cite{SSPP,hajimiri_dynamic_focusing,HashemiNature,JaffeJoM}, rapidly deployable infrastructure \cite{SSPParxiv,SSPPgdoutos}, aerodynamic vehicle microwave systems \cite{shape_austin}, and more \cite{suresh2022origami_sengupta}.

Collapsible \cite{fikes_cocured}, flexible \cite{fikes_shapecal_TMTT,mizrahi_shapecal_IMS}, inflatable \cite{huang_inflatable}, and foldable \cite{FIKES_JOM_2023} antenna array prototypes have been demonstrated. While these prototypes have shown they are mechanically and electrically capable of suiting the needs of the aforementioned applications, large-scale realization cannot come to fruition unless they prove to be economically efficient relative to the intended markets. To date, a techno-economic analysis projecting the cost of FPAs has not been published.

Presented in this paper is a cost model that seeks to fill this gap. Envisioned is a model FPA that is light, durable, low-cost, thin, and boasts functionalities packaged into an RFIC. The manufacturing process for such a FPA is discussed with respect to four ``layers:" the IC, the flexible PCB, the radiators, and assembly. At each phase, three tiers of cost projection are discussed: a ``current tier" which assumes no changes to the current manufacturing process, a ``middle tier" which assumes large scale manufacturing and conservative projections for the future, and an ``asymptotic" tier which assumes large scale manufacturing and great advancements in technology. The goal is to arrive at a projection for the per area cost of producing a FPA as herein envisioned, with a low and high bound coming from the lowest and highest projection tiers. We also discuss a mass projection and use capital expense estimates to project a startup costs.

While the envisioned FPA is well-suited for terrestrial and aquatic communications, wireless power, and sensing, our analysis is particularly targeted for applications in space, such as satellite communications and space solar power, where large aperture, low mass, and deployability are critical and where current solutions are very limited.

In Section II, the model FPA is discussed. In Section III, the manufacturing process is discussed and relevant cost details are presented. In Section IV, the total FPA sheet cost is presented. In In Section V, a total startup cost is presented. In Section VI, an aerial mass density projection is presented. Section VII, a cost comparison to various competitors in different markets is presented. Listed in the Appendix are cost model variables, and a discussion on price sensitivity to design changes.

\section{A Model Flexible Phased Array Sheet}

\begin{figure*}
	\centering
    \includegraphics[width=\textwidth]{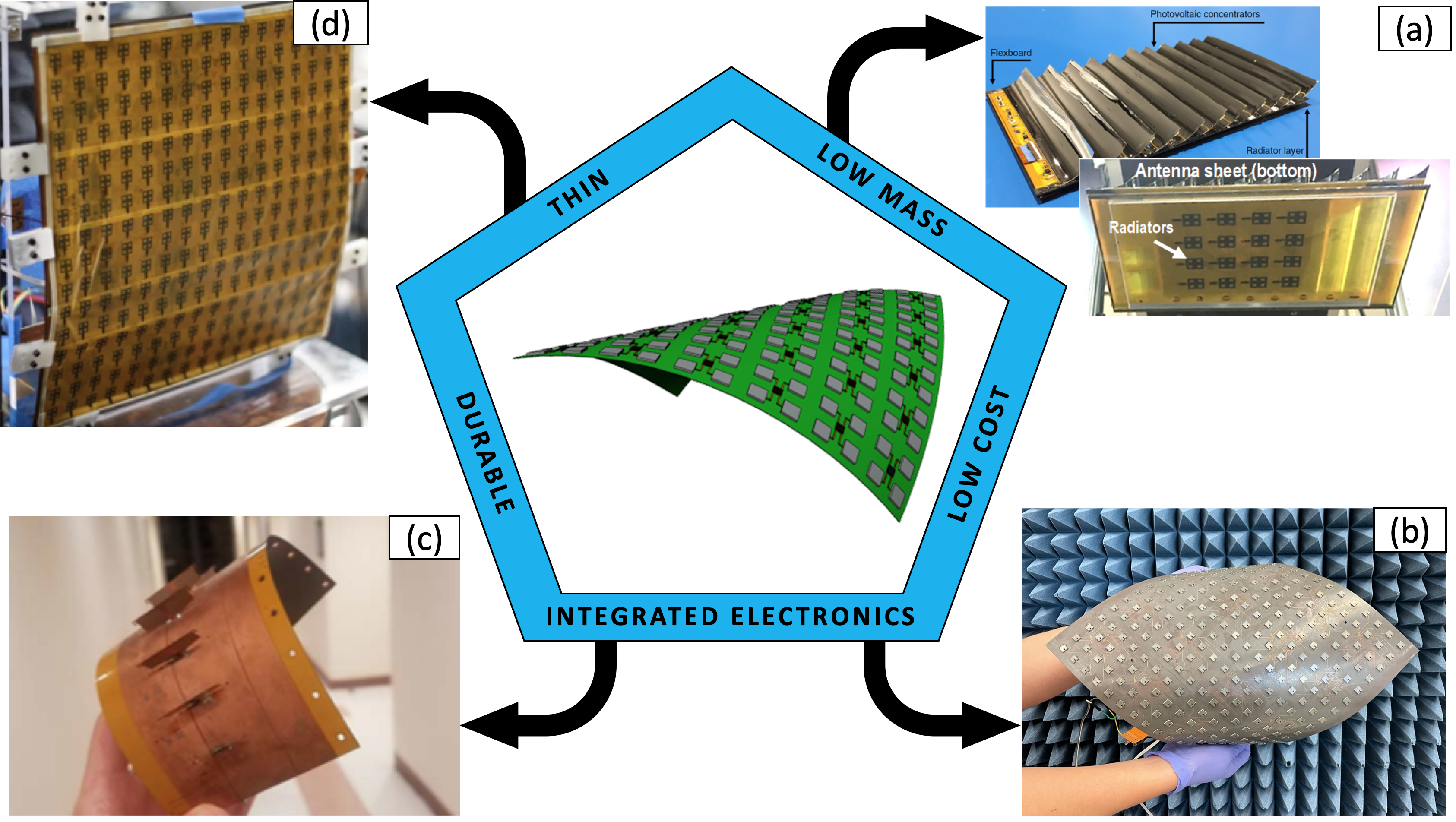}
	\caption{Center: Envisioning the ideal FPA: thin, low mass, durable, low cost, with integrated electronics. Outside, clockwise from the top right: (a) A wireless power transfer tile with photovoltaic concentrators on one side, a FPA in the middle, and 10GHz flexible patch antennas on the other side. (b) A 256-element FPA with integrated power generation and 10GHz patch antennas. (c) An 8-element FPA tile with 10GHz dipole radiators and integrated power control. (d) A 256-element FPA with integrated power generation and a flexible sheet of 10GHz patch antennas.}
	\label{Array_control_modular}
\end{figure*}

A model FPA can be stripped down to three fundamental components:
\begin{enumerate}
    \item A RFIC which is powered by a DC power supply and performs all the necessary frequency locking, power amplification, phase shifting, signal modulation, and digital control functionalities necessary for the desired application.
    \item A thin and flexible PCB with radiator ground plane, microstrip routing between the IC and the radiators, and microstrip lines for power and digital control.
    \item Thin and flexible radiators which are suited for the relevant physical context and application\footnote{Different radiators offer tradeoffs vis-a-vis radiation polarization, bandwidth, form-factor, etc. Almost all designs can be made flexible, as shown in various publications \cite{matan2022npj,fikes_cocured,FIKES_JOM_2023}.}.
\end{enumerate}

These descriptions are intentionally general: they are common to almost any imaginable embodiment of a FPA and variations on the details do not greatly affect the cost projection presented below. For example, because ICs are paid for by the wafer and not by the transistor, our model accounts for many applications that might use different IC architectures and building blocks (single frequency wireless power transfer, wideband communications, etc.), assuming area is not greatly affected.

However, to obtain some of the specific details presented herein, we make assumptions about the exact nature of the FPA:
\begin{enumerate}
    \item The RFIC is manufactured in 65nm CMOS, with an integrated micro-controller, bumping, and potentially an interposer.
    \item The PCB is etched out of 2-layer, rolled polyimide and copper clad.
    \item The radiators are 10GHz co-cured popup dipoles as presented in \cite{fikes_cocured}. The same composite-electromagnetic co-design can be applied to different radiator architectures with a similar cost.
\end{enumerate}

Other system variables used in the model are listed in the Appendix below:

\section{Cost Model}
\subsection{RFIC}
Fundamentally, the RFIC cost is a function of two numbers: The cost per wafer and the number of chips per wafer (wafer yield).
\subsubsection{Wafer Cost}
Assuming a 65 nm process, a 300mm wafer costs \$2000. Considering the additional \$60 for bumping the wafer, total wafer cost at scale, $C_\text{wafer}$, is 2060 \$/wafer.

\subsubsection{Wafer Yield}
Currently, the model 16-channel WPT RFIC \cite{HashemiNature} is 3mm x 3mm = 9mm$^2$. Adding an additional 0.1mm/side to account for the width of the dicing lane and an additional $\approx 4\text{mm}^2$ to integrated a microcontroller, total area is 13.61mm$^2$.

Wafer yield is not simply the quotient of the wafer area and the chip area; geometric incompatibility of cutting squares out of a circle reduces yield. True yield can be estimated using the following formula:
\begin{align}
    N_\text{chips} 
    &= \left\lfloor\frac{\pi R_\text{wafer}^2}{A_\text{chip}} - \frac{2\pi R_\text{wafer}}{\sqrt{2A_\text{chip}}}\right\rfloor\\
    &= \left\lfloor\frac{\pi \cdot (150 \text{mm})^2}{13.61\text{mm}^2} - \frac{2\pi \cdot 150\text{mm}}{\sqrt{2\cdot 13.61\text{mm}^2}}\right\rfloor\\
    &= \left\lfloor 5013.023 \right\rfloor\\
    &= 5013
\end{align}

Defect yield (the fraction of chips that pass inspection) is highly variable and subject to changes based on technology node, devices used, desired accuracy, etc. Assuming 5\% of chips are discarded because they contain defective devices, chip yield becomes 4762 chips per wafer. Thus, our chip cost ($C_\text{chip}$) is:
\begin{align}
    C_\text{chip} &= \frac{C_\text{wafer}}{N_\text{chips}}\\
                  &= \frac{\$ 2060}{4762}\\
                  &= \$0.4326/\text{chip}
\end{align}

Finally, price per unit area ($C_\text{chip}^A$) can be computed by dividing chip cost by the total area driven by a single chip (herein called a tile) ($A_\text{tile}$):
\begin{align}
    C_\text{chip}^A &= \frac{C_\text{chip}}{A_\text{tile}}\\
    &= \frac{C_\text{chip}}{Kd^2}\\
    &= \frac{\$ 0.4326}{16\cdot (18\times 10^{-2}\text{m})^2}\\
    &= 83.45 \frac{\$}{\text{m}^2}
\end{align}
where $K$ is the number of radiators per chip and $d = 0.6\lambda$ is the radiator spacing.

Additional costs include the cost to dice the wafer (assumed to be 3\% of the chip cost), the cost to produce the mask set (\$ 500,000 for 65nm, amortized over lifetime production), and the cost to make an interposer, which we assume is made of the same PCB (and carrying the same cost per unit area) as that discussed in the next section). If the interposer is 3x the area of the chip, our final cost per unit area for the RFIC is $88.39 \frac{\$}{\text{m}^2}$ - a marginal increase over the cost before these additional costs.

This cost projection is based on our current process, but if we account for an improved process and lower prices with further advancements in technology nodes, we can project both a ``middle" projection and an ``asymptotic" projection:

\begin{table}[!h]
  \centering
  \begin{tabular}{r|c|c|c|c}
  Variable         & Unit       & Current & Middle  & Asymptotic\\
        \hline
$C_\text{wafer}$ (65nm, 300mm) & [\$/wafer] & 2000    & 1500    & 1200\\
Bumping            & [\$/wafer] & 60      & 40      & 20\\
Mask set           & [\$]       & 500,000 & 350,000 & 250,000\\
RF chip area       & [mm$^2$]   & 9       & 6       & 4\\
$\mu$C chip area   & [mm$^2$]   & 4       & 2       & 1\\
Dicing Lane width  & [$\mu$m]   & 100     & 70      & 50\\
Interposer         & [ ]        & YES     & YES     & NO\\
\hline\hline
Total Price        & [\$/m$^2$] & 88.39   & 39.67   & 19.51
\end{tabular}
  \caption{RFIC Cost Projections}
  \label{tab:1}
\end{table}

\subsection{Thin, Flexible PCB}
Flexible PCBs are new and uncommon and, thus, their manufacturing process is less known - especially at high scale.
\subsubsection{Clad Manufacturing}
The process begins with the construction of the PCB clad: a stack of layers of interleaved metal and insulator. Assumed herein are thin layers of copper and polyimide (Kapton). Clad comes either rolled or autoclaved; rolled clad is about 6x cheaper (owing to higher throughput) but has higher surface roughness \footnote{The inner surface of the metal has a high density of dendrites which allow it to adhere to the substrate without baking the stackup. The dendrites degrade the performance of microstrip lines and other electromagnetic structures.}. Assumed for the purposes of this model is the use of rolled clad, due to the importance of low-cost and scalability and the limited added losses.

The rolled clad is drilled to produce via holes which are then plated. The board is patterned and etched on both sides to generate RF and DC traces. Next, the board is conformal coated to protect parts from the environment and from potential shorts. Higher tier projections assume the use of less conformal coating which reduces mass and cost.

The following table lists expenses tracked along this manufacturing process:
\begin{table}[!h]
  \centering
  \begin{tabular}{r|c|c|c|c}
  Variable   & Unit      & Current & Middle & Asymptotic\\
        \hline
Materials   & [\$/m$^2$] &   98.79 &   4.94 &   2.66\\
Labor       & [\$/m$^2$] &   14.25 &   1.78 &   1.78\\
Energy      & [\$/m$^2$] &    0.16 &   0.16 &   0.16\\
Capex       & [\$/m$^2$] &    1.11 &   1.11 &   1.11\\
\hline\hline
Total Price & [\$/m$^2$] &  114.31&    7.99 &   5.71
\end{tabular}
  \caption{Flexible PCB Cost Projections}
  \label{tab:1}
\end{table}
The ``current" number is dominated by materials and labor costs because it assumes clad is purchased already manufactured instead of rolled in-house; the other two categories add the cost of manufacturing clad but assume materials purchased are raw copper and polyimide sheets instead. The great disparity in materials cost is owing to a 40x markup of off-the-shelf clad relative to the raw materials.

\begin{figure*}[t]
\centering
 \includegraphics[width=\textwidth]{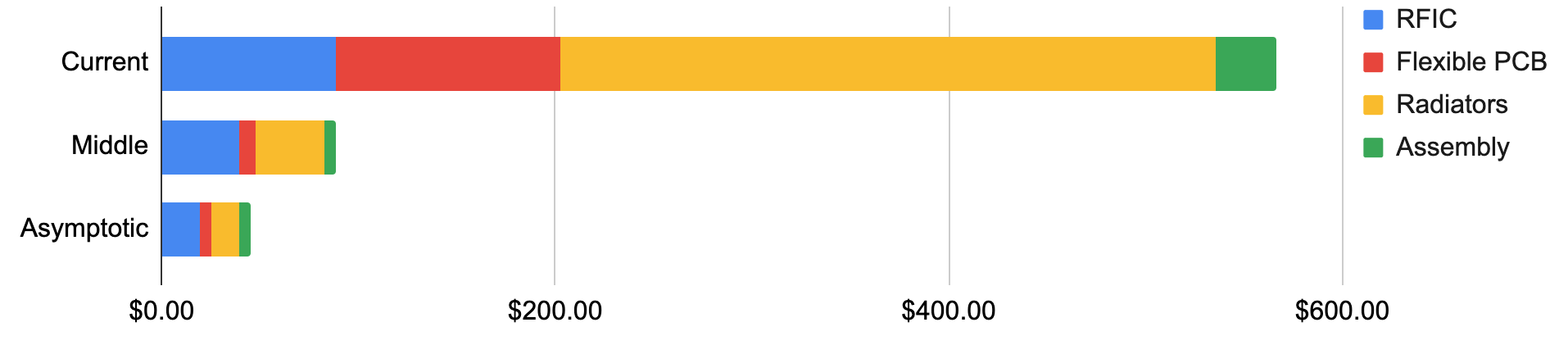}
\caption{FPA cost [\$/m$^2$] at different projection tiers: (a) Current (\$566.21/m$^2$), (b) Middle (\$88.75/m$^2$), and (c) Asymptotic (\$45.37/m$^2$).}
\label{samplingrate}
\end{figure*}

\subsection{Flexible Radiators}

Of critical importance in large, scalable, flexible, lightweight, and durable phased array systems are radiators which themselves have these characteristics. The most compelling platform for designing and manufacturing flexible radiators, presented in \cite{fikes_cocured}, is one in which an electronic antenna layer is ``co-cured" to a flexible substrate. This allows for complete control over the physical form, mechanical strength, and thermal properties of the radiator. Moreover, any incarnation of a flexible radiator using this platform would go through a similar manufacturing process, allowing for the projection of a wide range of radiators using a single model.

\subsubsection{Composite Substrate Preparation}
The manufacturing process begins with a single-ply composite mesh. The mesh can either be purchased pre-impregnated with an epoxy matrix or raw and then impregnated in house. Owing to the extremely high-cost of current off-the-shelf options, this model assumes in-house impregnation for the middle and asymptotic tiers.

The impregnated fiber plies are stacked with respect to a particular weave\footnote{Assumed in this case is the 3-ply 45$^\circ$/90$^\circ$/45$^\circ$ stack presented in \cite{fikes_cocured}}. The plies are laminated under light heat and cut into the final fiber shape.

\subsubsection{Electronic Antenna Preparation}
The electronic layer of the antenna - envisioned as the etched, 2-layer polyimide flexible PCB used in \cite{fikes_cocured} - is prepared in parallel. These antennas are processed in a manner similar to the PCB but without any conformal coating and utilizing a much thinner substrate. Furthermore, empty area on antenna sheets not containing metal traces are cut out to reduce mass and dielectric that might interfere electromagnetically.

\subsubsection{Alignment and Curing}
Finally, the fiber backing and the antenna layers are aligned and laminated to produce the final radiator stack. This stack is then fit into a custom silicone mold which is vacuum-sealed and cured under temperature and pressure for 2 hours. The combined curing of both layers - called ``co-curing" - is a method of producing flexible radiators with great potential in scalability without compromising on consistency.

Listed below are estimates of costs for the various stages of manufacturing:
\begin{table}[H]
  \centering
  \begin{tabular}{r|c|c|c|c}
  Variable & Unit & Current & Middle & Asymptotic\\
        \hline
Impregnated Fiber       & [\$/m$^2$] & 196.85 &  16.79 &   1.68\\
Polyimide clad          & [\$/m$^2$] &  86.00 &   0.53 &   0.53\\
Fiber/Curing Processing & [\$/m$^2$] &   5.02 &   5.02 &   5.02\\
Antenna Processing      & [\$/m$^2$] &  21.94 &   4.47 &   4.47\\
Labor                   & [\$/m$^2$] &   8.75 &   0.88 &   0.88\\
Capex                   & [\$/m$^2$] &  14.68 &   7.41 &   1.59\\
\hline\hline
Total Price             & [\$/m$^2$] & 333.24 &  35.09 &  14.16
\end{tabular}
  \caption{Co-Cured Popup Antenna Cost Projections}
  \label{tab:1}
\end{table}
The current projection is dominated by materials costs owing to the extremely high price of pre-impregnated fiber and off-the-shelf polyimide clad. When produced in house, these materials costs reduce by an order of magnitude. Further savings are achieved through increased automation and reduced capex by assuming more durable silicone molds in the latter tiers.

\subsection{Assembly}
Finally, the three components are assembled using highly scalable manufacturing steps that are extremely standard in industry and easy to model: pick-and-place assembly and flip-chip bonding for IC. In addition, the cost of several resistors, capacitors (``passives"), and a high emissivity metal ``thermal spreader" mounted on top of the RFIC is accounted for.
\begin{table}[!h]
  \centering
  \begin{tabular}{r|c|c|c|c}
  Variable   & Unit       & Current & Middle & Asymptotic\\
        \hline
Passives     & [\$/m$^2$] &  5.15 &  2.58 &  2.58\\
Processing   & [\$/m$^2$] &  0.25 &  0.21 &  0.21\\
Labor        & [\$/m$^2$] & 23.31 &  1.93 &  1.93\\
Capex        & [\$/m$^2$] &  1.55 &  1.29 &  1.29\\
\hline\hline
Total Price  & [\$/m$^2$] & 30.26 &  6.00 &  6.00
\end{tabular}
  \caption{WPT Layer Assembly Cost Projections}
  \label{tab:1}
\end{table}
In this case, cost reductions come mostly from high automation and redesigning the ICs to use fewer passive components.

\section{Flexible Phased Array Sheet Cost}

The full, per-area cost for the envisioned FPA sheet can be projected by summing the costs from the four manufacturing steps discussed above:
\begin{table}[!h]
  \centering
  \begin{tabular}{r|c|c|c|c}
  Variable   & Unit       & Current & Middle & Asymptotic\\
        \hline
RFIC         & [\$/m$^2$] &  89.23 &  39.67 &  19.51\\
Flexible PCB & [\$/m$^2$] & 114.31 &   7.99 &   5.71\\
Radiators    & [\$/m$^2$] & 333.24 &  35.09 &  14.16\\
Assembly     & [\$/m$^2$] &  30.26 &   6.00 &   6.00\\
\hline\hline
Total Price  & [\$/m$^2$] & 566.21 &  88.75 &  45.37
\end{tabular}
  \caption{FPA Sheet Cost Projection}
  \label{tab:1}
\end{table}

As we can see in the ``Middle" column, scaling up and automating current processes is responsible for a 6x reduction in cost. Moreover, the entire bound between current and asymptotic spans only a single order of magnitude, providing us with a narrow estimate for FPA sheet cost.

\begin{figure*}[t]
\centering
 \includegraphics[width=\textwidth]{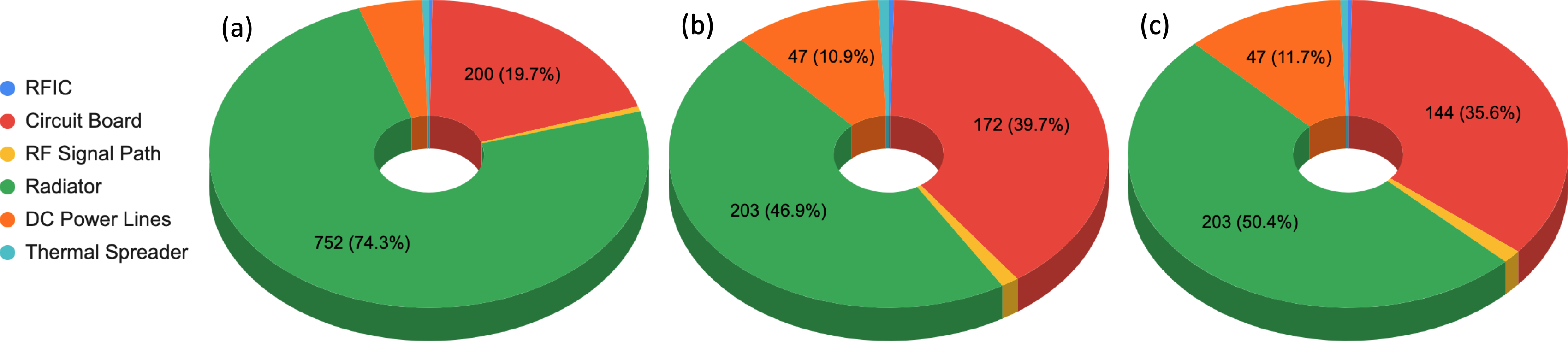}
\caption{Tile mass [mg] at different projection tiers: (a) Current (1012mg), (b) Middle (433mg), and (c) Asymptotic (403mg).}
\label{samplingrate}
\end{figure*}

\section{Startup Cost}
Relevant to large-scale production of an device with a novel manufacturing chain is the cost to outfit a facility for production. Startup cost, measured in $\frac{\$\cdot \text{yr}}{\text{m}^2}$, was projected by tracking only the capital expenses and normalizing them by their associated annual throughput, measured in m$^2/\text{yr}$. This number is subsequently scaled by the desired facility capacity, measured in m$^2/\text{yr}$, to calculate the cost of the facility.

This cost does not include the price of land or taxes, which are subject to changes based on location, market conditions, local taxes, and potential subsidies obtained from local or federal government(s). Moreover, these startup costs are not {\it in addition} to the total costs per unit array area presented earlier; per-unit-array-area costs include capital expenditures that overlap with those included in the startup cost, though many of the capital expenses will be recurring based on machine lifetime. Startup costs can be seen simply as representing the initial investment necessary before any arrays can be produced or sold.

\begin{table}[!h]
  \centering
  \begin{tabular}{r|c|c|c|c}
  Variable   & Unit       & Current & Middle & Asymptotic\\
        \hline
Flexible PCB & [$\$\cdot \text{yr}/\$\text{m}^2$] &  31.68 &  31.68 &  31.68\\
Radiators    & [$\$\cdot \text{yr}/\$\text{m}^2$] &  37.18 &  37.18 &  37.18\\
Assembly     & [$\$\cdot \text{yr}/\$\text{m}^2$] &  31.13 &  25.76 &  25.76\\
\hline\hline
Total Price  & [$\$\cdot \text{yr}/\$\text{m}^2$] &  99.98 &  94.61 &  94.61
\end{tabular}
  \caption{Flexible Phased Array Sheets Startup Cost}
  \label{tab:1}
\end{table}

Projections are similar for all tiers owing to equal machine pricing and roughly constant throughput. For a facility capable of producing 10,000 m$^2$ of FPA sheets in a year, an initial investment of between \$946,100-\$999,800 would be required.

\section{Aerial Mass Density}
A mass model is populated by investigating the same three manufacturing layers discussed above: each component contributes predictable mass. These masses are tabulated below to generate a projection of the aerial mass density of a single FPA sheet tile:

\begin{table}[H]
    \centering
    \begin{tabular}{r|c|c|c|c}
         Component          & Unit      & Current & Middle & Asymptotic\\
         \hline
         RFIC               & [mg]      &    2.55 &   1.70 &   1.13\\
         Flexible PCB       & [mg]      &  199.58 & 171.59 & 143.60 \\
         RF signal path     & [mg]      &    5.60 &   5.60 &   5.60\\
         Radiator           & [mg]      &  752.00 & 203.04 & 203.04\\
         DC power lines     & [mg]      &   47.24 &  47.24 &  47.24\\
         Thermal spreader   & [mg]      &    5.04 &   3.36 &   2.24\\
         \hline\hline
         Total mass         & [mg]      & 1012.01 & 432.53 & 402.85\\
         Total mass density & [g/m$^2$] &  195.22 &  83.43 &  77.71
    \end{tabular}
    \caption{FPA Tile Mass Projection}
    \label{tab:WPT_mass}
\end{table}

Mass reductions in the tiers can be attributed to design decisions that were also responsible for cost reductions. Higher tiers assume smaller RFICs. This also reduces the size (and thus mass) of the thermal spreader which is sized in proportion to the IC. Some mass reductions, like that for the flexible PCB and radiators, are owing to design decisions made specifically to reduce mass. The middle and asymptotic tiers assume ``cheesing"\footnote{``Cheesing" refers to the etching of a grid pattern in an otherwise solid ground plane at a pitch much lower than the wavelength for the purpose of minimizing mass.} in the PCB ground plane and less glass fiber substrate in the radiator. Some components, such as the DC power lines and RF signal path, are sized based on electrical considerations and won't reduce in mass with higher tiers.

\section{Market Competitiveness}
Absolute cost is useful but far more relevant is cost efficiency relative to market competitors. As there are no FPAs on the market, the closest competitors are other phased array solutions for the relevant applications. Currently, there are market offerings for typical phased array transceivers, but more relevant are the offerings advertising mobile connectivity (portable transceivers and arrays that are designed to be mounted on cars, boats, and planes).

The cost comparison below was enabled by the relentless gains in the number of offerings and in the market size. There are now consumer offerings (like Starlink's ``Dishy McFlatface") which have spurred detailed breakdowns and in-depth device specifications. Elusive details were filled in using assumptions listed below.

\begin{table*}[!h]
  \centering
  \begin{tabular}{r|c|c|c|c|c}
             &            & Starlink   & Starlink    & Kymeta & KVH \\
 Variable    & Unit       & (Standard) & (High Perf) & (u8)   & (TracVision A9)\\
        \hline
Electronics Cost\tablefootnote{Assumed to be 80\% of market cost.} 
             & [\$]       &    479 &  2,000 & 15,160 & 46,800\\
Electronics Mass\tablefootnote{Assumed to be 25\% of total mass if number wasn't available. 25\% is based on the measured ratio for Starlink's Dishy McFlatface V2.}
             & [g]        &    725 &  1,725 &  8,000 & 5,625\\
Area         & [m$^2$]    &  0.155 &  0.294 &  0.810 & 0.656\\
\hline\hline
Cost density & [\$/m$^2$] &   3,083 &   6,807 & 18,716 & 7,316\\
Mass density & [g/m$^2$]  &   4,664 &   5,870 &  9,877 & 8,573
\end{tabular}
  \caption{Phased Array Cost and Mass Comparison Table}
  \label{tab:1}
\end{table*}

The best comparison on the market is Starlink's ``Standard" transceiver, because it is being produced at scale with among the most modern hardware on the market, which has aerial mass density of 4,664 g/m$^2$ and a cost density of 3,083 \$/m$^2$. Even adjusting the presented cost model to a frequency of 40GHz (in the middle of Starlink's uplink and downlink bands), this model projects an aerial mass density of 2,396 g/m$^2$ and a cost density of 761 \$/m$^2$ (for the ``middle" projection).

% other specific markets - what are our competitors - what is pricing for those?\\
% starlink? qualcomm arrays? 5G?

% space-born foldable arrays -
% Conformal array solutions? - military planes with phased arrays?
% communications in transport - yachts - metamaterial phased arrays - Kymeta??

% TRY to find other benchmark prices that we can compare to???

% https://ieeexplore.ieee.org/stamp/stamp.jsp?tp=&arnumber=7350112
% 183,312/m^2 for a tile design (2015)

\section{Conclusion}

Presented here is an envisioned model FPA and a corresponding cost model which projects the price per unit area of producing it at large scale. The FPA is constructed of an RFIC and flexible radiators mounted to a thin, flexible PCB. This platform is equipped for numerous applications in communications, sensing, and wireless power transfer; small changes to design details do not greatly affect the projected cost.

The cost model projects prices at three tiers, with each tier making progressively more assumptions about scaling and future technology pricing. Using the ``current" projection which assumes current pricing and no scaling, FPAs can be produced at \$566.21 per square meter, a number strongly inflated by off-the-shelf materials costs. More accurate for a high-scale production is the ``middle" projection, which estimates production at \$88.75 per square meter. With further advances in technology, the asymptotic price limit is projected at \$45.37 per square meter.

Presented alongside the cost model is a mass model which projects an aerial mass density of between 195.22 g/m$^2$ and 77.71 g/m$^2$ for the current and asymptotic tiers, respectively. Additionally, capex costs throughout the model were collected and normalized to annual throughput to generate a startup cost rate for the desired facility capacity. A facility producing 10,000 m$^2$ of FPA sheets would require an up-front investment of about \$1 million.

Projected prices are highly competitive relative to current offerings in the market and are approximately $1/4$ the cost of the leading rigid phased array transceiver. Projected aerial mass density is also between one-half and one-tenth as large as current rigid offerings. The envisioned FPA can enable a factor reduction in launch costs for satellites and can be integrated into air-borne designs in ways current market offerings are simply incapable of.

This analysis lends credence to the notion that FPAs are not only mechanically and electrically qualified to be implemented in the desired applications, but also economically efficient relative to the corresponding markets. FPAs are relatively cheap to produce, extremely light, and require a small initial capital investment to produce. Low-cost FPAs are uniquely positioned to enable the deployment of ultra-large apertures in novel contexts - space, aquatic and aerodynamic vehicles, extreme conditions, and others - facilitating greater power, higher bandwidth, higher power levels, and new applications previously prohibited or not yet envisioned.

\section{Appendix}
\subsection{System Variables}
Listed below are variables assumed for this model.
\begin{table}[H]
  \centering
    \begin{tabular}{c|c|c|c}
     Symbol & Description               & Unit     & Value\\
     \hline
     $f$    & Frequency                 & [GHz]    & 10\\
     $d$    & Radiator spacing          & [mm]     & 18 ($0.6\lambda$)\\
     $K$    & Output channels per RFIC  & [ ]      & 16\\
     $L$    & Flexible PCB metal layers & [ ]      & 2\\
     $t_m$  & Metal thickness           & [$\mu$m] & 10\\
     $t_p$  & Polyimide thickness       & [$\mu$m] & 12.5
\end{tabular}
  \caption{FPA Variables and Assumed Values}
  \label{tab:1}
\end{table}

\subsection{Economic Variables}

A few market variables factored into our calculations. These are middle values that change as a function of time and place but were chosen to represent realistic market conditions:
\begin{table}[H]
    \centering
    \begin{tabular}{c|c|c}
    Variable Description & Unit & Value\\
    \hline
     Cost of labor, adjusted\tablefootnote{Adjustment is based on a cost of wage of \$25/hour adjusted by 2x to include the cost of benefits, workplace insurance, labor expenses (cleaning, bathroom, facilities, etc.), and other expenses.} & [\$/hour] & 50 \\
     Cost of electricity\tablefootnote{In Texas, USA. 2022} & [\textcent/kWh] & 8.05\\
     Inflation rate & [\%/year] & 3.8
    \end{tabular}
    \caption{Economic Variables and Assumed Values}
    \label{tab:my_label}
\end{table}

% www.energybot.com/electricity-rates/texas/\#:~:text=The\%20average\%20Texas\%20commercial\%20electricity,lower\%20than\%20the\%20national\%20average).

\subsection{Cost Variability With Different Design Variables}

Though some design decisions were necessary to identify certain prices, the model is fairly amenable to changes in these details. We used this flexibility to demonstrate how different design variables would affect the final cost. All percentages are relative to the ``middle" projection of 88.75 \$/m$^2$.

\begin{itemize}
    \item Changing from a Si CMOS process to a GaAs process increases the IC cost by a factor of 4.8x. This increases total cost to 239.50 \$/m$^2$ (a 170\% increase).
    \item Increasing operating frequency increases the density of RFICs, increasing overall cost. At $f = 20$GHz, total cost is 220.62 \$/m$^2$ (a 149\% increase); at $f = 40$GHz, total cost is 761.35 \$/m$^2$ (a 758\% increase).
    \item Increasing the number of PCB layers from 2 to 4, for the purpose of more real-estate for routing or advanced functionalities, increases cost slightly to 98.40 \$/m$^2$ (a 10\% increase).
    \item Autoclaved clad, necessary for applications requiring extremely low surface roughness for low attenuation, increases total cost to 113.00 \$/m$^2$ (a 28\% increase).
    \item Adding a conformal coating to protect the electronics in harsh environments (outdoors, high radiation conditions, space, etc.) increases total cost to 91.60 \$/m$^2$ (a 3\% increase).
\end{itemize}

\bibliographystyle{IEEEtran}
\bibliography{MyRef.bib}

\begin{thebibliography}{10}
\providecommand{\url}[1]{#1}
\csname url@samestyle\endcsname
\providecommand{\newblock}{\relax}
\providecommand{\bibinfo}[2]{#2}
\providecommand{\BIBentrySTDinterwordspacing}{\spaceskip=0pt\relax}
\providecommand{\BIBentryALTinterwordstretchfactor}{4}
\providecommand{\BIBentryALTinterwordspacing}{\spaceskip=\fontdimen2\font plus
\BIBentryALTinterwordstretchfactor\fontdimen3\font minus
  \fontdimen4\font\relax}
\providecommand{\BIBforeignlanguage}[2]{{%
\expandafter\ifx\csname l@#1\endcsname\relax
\typeout{** WARNING: IEEEtran.bst: No hyphenation pattern has been}%
\typeout{** loaded for the language `#1'. Using the pattern for}%
\typeout{** the default language instead.}%
\else
\language=\csname l@#1\endcsname
\fi
#2}}
\providecommand{\BIBdecl}{\relax}
\BIBdecl

\bibitem{YangMTT}
B.~{Yang}, Z.~{Yu}, J.~{Lan}, R.~{Zhang}, J.~{Zhou}, and W.~{Hong}, ``Digital
  beamforming-based massive mimo transceiver for 5g millimeter-wave
  communications,'' \emph{IEEE Transactions on Microwave Theory and
  Techniques}, vol.~66, no.~7, pp. 3403--3418, 2018.

\bibitem{Natarajan77GHzTX2006}
A.~Natarajan, A.~Komijani, X.~Guan, A.~Babakhani, and A.~Hajimiri, ``A 77-ghz
  phased-array transceiver with on-chip antennas in silicon: Transmitter and
  local lo-path phase shifting,'' \emph{IEEE Journal of Solid-State Circuits},
  no.~12, pp. 2807--2819, Dec. 2006.

\bibitem{FatemiOpt}
\BIBentryALTinterwordspacing
R.~Fatemi, B.~Abiri, A.~Khachaturian, and A.~Hajimiri, ``High sensitivity
  active flat optics optical phased array receiver with a two-dimensional
  aperture,'' \emph{Opt. Express}, vol.~26, no.~23, pp. 29\,983--29\,999, Nov
  ts. [Online]. Available:
  \url{http://www.opticsexpress.org/abstract.cfm?URI=oe-26-23-29983}
\BIBentrySTDinterwordspacing

\bibitem{McIntoshGRS}
R.~E. {McIntosh}, S.~J. {Frasier}, and J.~B. {Mead}, ``Fopair: a focused array
  imaging radar for ocean remote sensing,'' \emph{IEEE Transactions on
  Geoscience and Remote Sensing}, vol.~33, no.~1, pp. 115--124, 1995.

\bibitem{BrautigamGRS}
B.~{Brautigam}, J.~H. {Gonzalez}, M.~{Schwerdt}, and M.~{Bachmann},
  ``Terrasar-x instrument calibration results and extension for tandem-x,''
  \emph{IEEE Transactions on Geoscience and Remote Sensing}, vol.~48, no.~2,
  pp. 702--715, 2010.

\bibitem{MatanIMS2020}
M.~{Gal-Katziri}, A.~{Fikes}, F.~{Bohn}, B.~{Abiri}, M.~R. {Hashemi}, and
  A.~{Hajimiri}, ``Scalable, deployable, flexible phased array sheets,'' in
  \emph{2020 IEEE/MTT-S International Microwave Symposium (IMS)}, 2020, pp.
  1085--1088.

\bibitem{matan2022npj}
M.~Gal-Katziri, A.~Fikes, and A.~Hajimiri, ``Flexible active antenna arrays,''
  \emph{{npj Flex Electron.}}, vol.~6, no.~85, October 2022.

\bibitem{IMS_2022_A}
K.~Hu, G.~Soto–Valle, Y.~Cui, and M.~M. Tentzeris, ``Flexible and scalable
  additively manufactured tile-based phased arrays for satellite communication
  and 50 mm wave applications,'' in \emph{2022 IEEE/MTT-S International
  Microwave Symposium - IMS 2022}, 2022, pp. 691--694.

\bibitem{IMS_2022_B}
X.~Wang, D.~You, X.~Fu, H.~Lee, Z.~Li, D.~Awaji, J.~Pang, A.~Shirane,
  H.~Sakamoto, and K.~Okada, ``A flexible implementation of ka-band active
  phased array for satellite communication,'' in \emph{2022 IEEE/MTT-S
  International Microwave Symposium - IMS 2022}, 2022, pp. 753--756.

\bibitem{SSPP}
A.~Fikes, M.~Gal-Katziri, E.~Gdoutos, M.~Kelzenberg, E.~Warmann, R.~Madonna,
  H.~Atwater, A.~Hajimiri, and S.~Pellegrino, ``The caltech space solar power
  project: Design, progress, and future direction,'' in \emph{IEEE WiSEE Space
  Solar Power Workshop}, 2022.

\bibitem{hajimiri_dynamic_focusing}
A.~Hajimiri, B.~Abiri, F.~Bohn, M.~Gal-Katziri, and M.~H. Manohara, ``Dynamic
  focusing of large arrays for wireless power transfer and beyond,'' \emph{IEEE
  Journal of Solid-State Circuits}, vol.~56, no.~7, pp. 2077--2101, 2021.

\bibitem{HashemiNature}
\BIBentryALTinterwordspacing
M.~R.~M. Hashemi \emph{et~al.}, ``A flexible phased array system with low areal
  mass density,'' \emph{Nature Electronics}, vol.~2, no.~5, pp. 195--205, May
  2019. [Online]. Available: \url{https://doi.org/10.1038/s41928-019-0247-9}
\BIBentrySTDinterwordspacing

\bibitem{JaffeJoM}
C.~T. Rodenbeck, P.~I. Jaffe, B.~H. Strassner~II, P.~E. Hausgen, J.~O.
  McSpadden, H.~Kazemi, N.~Shinohara, B.~B. Tierney, C.~B. DePuma, and A.~P.
  Self, ``Microwave and millimeter wave power beaming,'' \emph{IEEE Journal of
  Microwaves}, vol.~1, no.~1, pp. 229--259, 2021.

\bibitem{SSPParxiv}
\BIBentryALTinterwordspacing
B.~Abiri, M.~Arya, F.~Bohn, A.~Fikes, M.~Gal-Katziri, E.~Gdoutos, A.~Goel,
  P.~E. Gonzalez, M.~Kelzenberg, N.~Lee, M.~A. Marshall, T.~Roy, F.~Royer,
  E.~C. Warmann, T.~Vinogradova, R.~Madonna, H.~Atwater, A.~Hajimiri, and
  S.~Pellegrino, ``A lightweight space-based solar power generation and
  transmission satellite,'' 2022. [Online]. Available:
  \url{https://arxiv.org/abs/2206.08373}
\BIBentrySTDinterwordspacing

\bibitem{SSPPgdoutos}
\BIBentryALTinterwordspacing
E.~Gdoutos, C.~Leclerc, F.~Royer, M.~D. Kelzenberg, E.~C. Warmann,
  P.~Espinet-Gonzalez, N.~Vaidya, F.~Bohn, B.~Abiri, M.~R. Hashemi,
  M.~Gal-Katziri, A.~Fikes, H.~Atwater, A.~Hajimiri, and S.~Pellegrino, \emph{A
  lightweight tile structure integrating photovoltaic conversion and RF power
  transfer for space solar power applications}. [Online]. Available:
  \url{https://arc.aiaa.org/doi/abs/10.2514/6.2018-2202}
\BIBentrySTDinterwordspacing

\bibitem{shape_austin}
A.~{Fikes}, A.~{Safaripour}, F.~{Bohn}, B.~{Abiri}, and A.~{Hajimiri},
  ``Flexible, conformal phased arrays with dynamic array shape
  self-calibration,'' in \emph{2019 IEEE IMS}, 2019, pp. 1458--1461.

\bibitem{suresh2022origami_sengupta}
\BIBentryALTinterwordspacing
S.~Venkatesh, D.~Sturm, X.~Lu, R.~J. Lang, and K.~Sengupta, ``Origami microwave
  imaging array: Metasurface tiles on a shape-morphing surface for
  reconfigurable computational imaging,'' \emph{Advanced Science}, vol.~9,
  no.~28, p. 2105016, 2022. [Online]. Available:
  \url{https://onlinelibrary.wiley.com/doi/abs/10.1002/advs.202105016}
\BIBentrySTDinterwordspacing

\bibitem{fikes_cocured}
A.~Fikes, O.~S. Mizrahi, A.~Truong, F.~Wiesemüller, S.~Pellegrino, and
  A.~Hajimiri, ``Fully collapsible lightweight dipole antennas,'' in \emph{2021
  IEEE International Symposium on Antennas and Propagation and USNC-URSI Radio
  Science Meeting (APS/URSI)}, 2021, pp. 545--546.

\bibitem{fikes_shapecal_TMTT}
A.~Fikes, O.~S. Mizrahi, and A.~Hajimiri, ``A framework for array shape
  reconstruction through mutual coupling,'' \emph{IEEE Transactions on
  Microwave Theory and Techniques}, vol.~69, no.~10, pp. 4422--4436, 2021.

\bibitem{mizrahi_shapecal_IMS}
O.~S. Mizrahi, A.~Fikes, and A.~Hajimiri, ``Flexible phased array shape
  reconstruction,'' in \emph{2021 IEEE MTT-S International Microwave Symposium
  (IMS)}, 2021, pp. 31--33.

\bibitem{huang_inflatable}
J.~Huang, ``The development of inflatable array antennas,'' \emph{IEEE Antennas
  and Propagation Magazine}, vol.~43, no.~4, pp. 44--50, 2001.

\bibitem{FIKES_JOM_2023}
A.~C. Fikes, M.~Gal-Katziri, O.~S. Mizrahi, D.~E. Williams, and A.~Hajimiri,
  ``Frontiers in flexible and shape-changing arrays,'' \emph{IEEE Journal of
  Microwaves}, vol.~3, no.~1, pp. 349--367, 2023.

\end{thebibliography}

\end{document}